\documentclass[superscriptaddress,preprint,aps,prl]{revtex4-1}

\usepackage{amssymb, amsmath}
\usepackage{graphicx}
\usepackage{dcolumn}
\usepackage{bm}

\newcommand{\bfo}[0]{BiFeO$_{3}$}
\newcommand{\lsmo}[0]{La$_{0.67}$Sr$_{0.33}$MnO$_{3}$}
\newcommand{\sto}[0]{SrTiO$_3$}

\begin{document}

\title{Element-Specific Depth Profile of Magnetism and Stoichiometry at the
       La$_{0.67}$Sr$_{0.33}$MnO$_{3}$/BiFeO$_3$ Interface}

\author{J.~Bertinshaw}
\author{S.~Br\"{u}ck}
\affiliation{School of Physics, University of New South Wales, Sydney, NSW 2052, Australia}
\affiliation{Australian Nuclear Science and Technology Organisation, Lucas Heights, NSW 2234, Australia}
\author{D.~Lott}
\affiliation{Institute for Materials Research, Helmholtz Zentrum Geesthacht, 21502 Geesthacht, Germany}
\author{H.~Fritzsche}
\affiliation{Canadian Neutron Beam Centre, Chalk River Laboratories, Ontario K0J 1J0, Canada}
\author{Y.~Khaydukov}
\author{O.~Soltwedel}
\author{T.~Keller}
\affiliation{Max-Planck-Institute for Solid State Research, outstation FRM II, D-70569 Stuttgart, Germany}
\author{E.~Goering}
\author{P.~Audehm}
\affiliation{Max-Planck-Institute for Intelligent Systems, D-70569 Stuttgart, Germany}
\author{D.~L.~Cortie}
\affiliation{Australian Nuclear Science and Technology Organisation, Lucas Heights, NSW 2234, Australia}
\author{W.~D.~Hutchison}
\affiliation{School of Physical, Environmental and Mathematical Sciences, University of New South Wales,
             Canberra, ACT 2600, Australia}
\author{Q. M. Ramasse}
\affiliation{SuperSTEM Laboratory, STFC Daresbury Campus, Keckwick Lane, Daresbury WA4 4AD, UK}
\author{M. Arredondo}
\affiliation{School of Mathematics and Physics, Queen's University Belfast, Belfast BT7 1NN, UK}
\author{R.~Maran}
\author{V.~Nagarajan}
\affiliation{School of Materials Science and Engineering, University of New South Wales, Sydney, NSW 2052, Australia}
\author{F.~Klose}
\affiliation{Australian Nuclear Science and Technology Organisation, Lucas Heights, NSW 2234, Australia}
\affiliation{Department of Physics and Materials Science, City University of Hong Kong, Hong Kong, SAR China}
\author{C.~Ulrich}
\affiliation{School of Physics, University of New South Wales, Sydney, NSW 2052, Australia}
\affiliation{Australian Nuclear Science and Technology Organisation, Lucas Heights, NSW 2234, Australia}

\date{\today}

\begin{abstract}
Depth-sensitive magnetic, structural and chemical characterization is important in the understanding and
optimization of novel physical phenomena emerging at interfaces of transition metal oxide heterostructures.
In a simultaneous approach we have used polarized neutron and resonant X-ray reflectometry to determine the
magnetic profile across atomically sharp interfaces of ferromagnetic \lsmo \slash multiferroic \bfo \ bi-layers
with sub-nanometer resolution. In particular, the X-ray resonant magnetic reflectivity measurements at the Fe and
Mn resonance edges allowed us to determine the element specific depth profile of the ferromagnetic moments in both
the \lsmo \ and \bfo \ layers. Our measurements indicate a magnetically diluted interface layer within the \lsmo \ layer,
in contrast to previous observations on inversely deposited layers \cite{Yu10}. Additional resonant X-ray reflection
measurements indicate a region of an altered Mn- and O-content at the interface, with a thickness matching that of
the magnetic diluted layer, as origin of the reduction of the magnetic moment.
\end{abstract}


\maketitle

Novel electronic states in transition metal oxide (TMO) thin film multilayered structures have created significant
attention recently~\cite{Yu10,Ohto04,Yama04,Reyr07,Mann10,Hwan12}. Unexpected properties, not present in the respective
bulk constituents, were demonstrated such as metallic conductivity at the interface between two insulators, or even
superconducting behavior~\cite{Reyr07}. Fundamentally, these states are a consequence of the symmetry breaking at
the interface between dissimilar oxide materials~\cite{Hwan12}. The resulting interface-near electronic states are defined
by spin exchange correlations, orbital reconstructions, band bending, Coulombic, magnetic, or superconducting penetration
into the adjacent layer, and epitaxial strain across the interface~\cite{Mann10,Hwan12,Chak07,Driz12}. As pointed out by
H.Y. Hwang $et$ $al.$ \cite{Hwan12}, their detailed origin is still under debate since standard experimental techniques do
not allow for the separation of intrinsic interface effects from modifications in the chemical composition, in particular
oxygen deficiencies. This requires new experimental techniques, which in particular can determine the chemical depth
profile across the interfaces with sub-nanometer resolution.

Two particularly interesting TMO materials are the multiferroic Bismuth Ferrite, BiFeO$_3$ (BFO), and the ferromagnetic
(FM) half-metal La$_{0.67}$Sr$_{0.33}$MnO$_3$ (LSMO), which constitutes an almost completely spin polarized electron system.
While for bulk LSMO the surface depolarizes the spins~\cite{Kono04} the problem can be largely overcome in TMO thin film
structures. BFO in itself is an exciting multiferroic compound, since it exhibits spontaneous magnetic (T$_N$ = 643~K) and
electric (T$_C$ = 1143~K) polarization well above room temperature~\cite{Cata09}. Both order parameters are indirectly coupled,
i.e. the spin state can be controlled through an electric field~\cite{Zhao06}. LSMO/BFO/LSMO trilayers exhibit a large
tunnel magnetoresistance (TMR). Interestingly, voltage-induced changes in the TMR behavior were recently demonstrated
in refs.~\cite{Bea08a,Hamb10}. This opens possibilities for multi-level memory state applications for next-generation information
storage devices~\cite{Eere06}. It is unclear, however, if the observed TMR effect is at its optimum. Any magnetic interface
modification leading to a perturbation of the spin-polarized flow at the LSMO interface would decrease the efficiency of
such a device~\cite{Yama04}. Therefore it is imperative to precisely identify the chemical and magnetic properties of the
interface states.

In this letter we demonstrate the influence of changes in the chemical composition on the interface magnetism
of LSMO\slash BFO bi-layers. The combination of complementary X-ray and neutron techniques provided a new and
unique insight into the magnetic and chemical depth profile of the bilayer interface, not accessible with standard
techniques. The simultaneous analysis of polarized neutron reflectivity (PNR) and element specific X-ray
resonant magnetic reflectometry (XRMR) provided a means of accurately determining the magnetic properties with
sub-nanometer resolution across the interface. Element specific resonant soft X-ray reflectivity (XRR) and XRMR
measurements performed at the Fe L$_{2,3}$ and Mn L$_{2,3}$-edges allowed for the determination of the magnetic
moments in the individual layers and indicated an altered stoichiometry at the interface as the reason for an
observed reduction of the magnetic moment.

LSMO (300$\,$\AA)\slash BFO (200$\,$\AA \ and 300$\,$\AA) bi-layers were grown epitaxially on a \sto\ (STO) (001) substrate
using a Neocera Pulsed laser Deposition (PLD) system with a $248\,\mathrm{nm}$ wavelength KrF excimer Laser. The atomic
precision of the layer-by-layer growth was calibrated by RHEED (reflection high electron energy diffraction). Oxygen
partial pressures of $100/10$\,$\mathrm{mTorr}$ and substrate temperatures of $900/850$\,$\mathrm{^\circ C}$ were held for
the deposition of the LSMO\slash BFO layers, respectively. The deposited bi-layer was cooled in a partial oxygen pressure
of $200$\,$\mathrm{Torr}$ at a rate of $20$\,$\mathrm{K/min}$ (for further details see ref.~\onlinecite{Hamb10}).
The bi-layers were initially characterized by laboratory X-ray diffraction and X-ray reflection (instrument X'Pert Pro),
which reveal an epitaxial growth with a root-mean squared (RMS) interface roughness of 5(1)$\,$\AA. This result is
consistent with scanning transmission electron microscopy (STEM) images on samples grown under identical conditions,
which show a sharp coherent interface (see Supplemental Material). These results demonstrate the excellent quality of our films,
and rules out interdiffusion larger than 5$\,$\AA\ across the interface. To compensate for in-plane stress caused
by the lattice mismatch, the c-axis lattice parameter was elongated in the $200$\,$\mathrm{\AA}$ thick BFO layer from
3.965\AA \ to 4.081(5)\AA, i.e. by 3.0~\%, and compressed for LSMO from 3.871$\,$\AA \ to 3.850(5)$\,$\AA, i.e. -0.55~\%.
This indicates a partial back-relaxation from the expected change in the c-axis lattice parameter, as observed in ref. \cite{Ange04}.

Magnetization measurements performed using Quantum Design PPMS and MPMS systems revealed that the LSMO layer orders
ferromagnetically below $\mathrm{T_C}=345(5)\,\mathrm{K}$ (see Supplemental Material). This corresponds to a
reduction of the expected bulk value of $370\,\mathrm{K}$ \cite{Urus95}. In-plane hysteresis curves collected
at $\mathrm{T}=150\,\mathrm{K}$ showed a low coercivity of $\sim$\,$1.7\,\mathrm{mT}$ and a saturated magnetic moment
of $3.0\,\mu_B$/Mn-ion, which is also smaller than the bulk values of $3.5\,\mu_B$/Mn-ion~\cite{Urus95}.
This is in accordance to previous experiments \cite{Ange04,Leco96,Koni99} and is primarily caused by the in-plane
epitaxial strain. The change in the c-axis lattice parameter, i.e. a modified c/a ratio, results in a biaxial tilt and
deformation of the MnO$_6$ octahedra, which affects the orbital arrangement. This weakens the in-plane hopping integral
and reduces the ferromagnetic double exchange interaction. No further effects, such as modified Mn- or O-content in the
bulk of the LSMO layer, are required to be accounted for, demonstrating the excellent quality of the films.

\begin{figure}
\centering
\includegraphics[width=6.8cm]{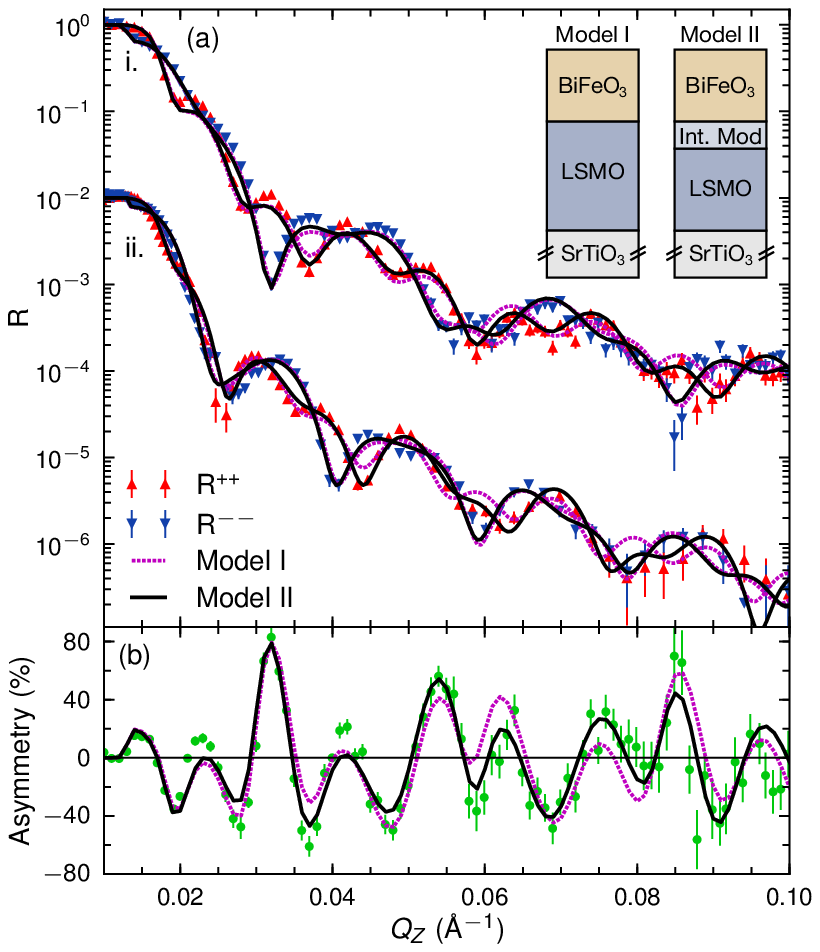}
\caption{
\label{fig:pnr_nrex_ref}
(color online)
(a) Specular PNR reflectivity for the $R^{++}$ and $R^{--}$ channels of a (i) 300$\,$\AA \ LSMO/200$\,$\AA \ BFO
sample taken at $\mathrm{T}=150\,\mathrm{K}$ on the instrument NREX and
(ii) of a 300$\,$\AA \ LSMO/300$\,$\AA \ BFO sample taken at $\mathrm{T}=300\,\mathrm{K}$ on the instrument Platypus
(the data are shifted in intensity).
(b) Asymmetry between both channels of experiment (i). The dashed lines depict the less accurate two layer simulation
without any interfacial layers (Model I). The solid lines correspond to the fit of the final model (Model II),
which is characterized by a $26\,$\AA \ interface layer with a $40\,\%$ suppressed magnetic moment.}
\end{figure}

Polarized Neutron Reflectometry is a powerful technique for the investigation of magnetic thin film systems \cite{Bea08}.
An accurate depth profile of the absolute ferromagnetic moment can be determined through changes in the specular
reflectance between spin-polarized neutron beams~\cite{Fitz04}. Reflectivity curves were measured using the instruments NREX,
located at FRM-II, Munich, Germany, D3 at NRU, Chalk River, Canada, and on Platypus at ANSTO, Australia. Figure
\ref{fig:pnr_nrex_ref} shows characteristic reflectivity data taken with a Q$_z$-resolution of $\sim$$4\%$.
The spin-up ($R^{++}$) and spin-down ($R^{--}$) neutron spin channels were measured to investigate the magnetic moment
of the bi-layer sample aligned parallel to an external applied field. Additional scans of the spin-flip channels $R^{+-}$
and $R^{-+}$, which probe the magnetic moment perpendicular to the neutron polarization, showed no measurable signal,
indicating that the LSMO magnetization was fully aligned within the plane of the film. Fig.~\ref{fig:pnr_nrex_ref}a)
shows the $R^{++}$ and $R^{--}$ spin-dependent specular reflectivities of two different bi-layer samples measured at
150$\,$K (i) and 300$\,$K (ii). The spin asymmetry ($A = \frac{R^{++} - R^{--}}{R^{++} + R^{--}}$) for the
150$\,$K data is plotted in Fig.~\ref{fig:pnr_nrex_ref}b). The temperature of 150$\,$K was chosen since it is
well below $\mathrm{T_C}$ of the LSMO layer and above a structural phase transition of the STO substrate at 110$\,$K
\cite{Shir69}. To ensure full saturation of the Mn-ion moments, the sample was field-cooled under an applied
magnetic field of $0.5\,\mathrm{T}$ parallel to the neutron polarization. Measurements were performed with an external
magnetic field of $0.7\,\mathrm{mT}$, required to maintain the polarization of the neutron beam. In this way, the effect
of an external magnetic field on the interface was minimized while maximizing the contrast between spin channels. The
SimulReflec software package was used to perform a simultaneous least squares fit \cite{SimulR}. Specific parameters
already determined by XRR, i.e. layer thicknesses, density and non-magnetic interface roughness, were fixed during the
fitting process. The dashed lines in Fig.~\ref{fig:pnr_nrex_ref} correspond to the fitting of a two-layer system of
LSMO and BFO without an interfacial layer (Model I). A significant improvement to the fit, with a reduction of the
residual error from 6.094 to 3.147 (see Supplemental Material), was obtained when including an additional magnetically
diluted interface layer of 26(5)$\,$\AA~with a $40\,\%$ suppressed magnetic moment of 1.8(2)$\,\mu_B$/Mn-ion instead of
the 2.6(3)$\,\mu_B$/Mn-ion of the bulk of the LSMO layer (Model II). Additional magnetic layers at the STO/LSMO interface
or at the BFO/air interface did not yield any convincing improvement of the fit.

\begin{figure}
\centering
\includegraphics[width=6.7cm]{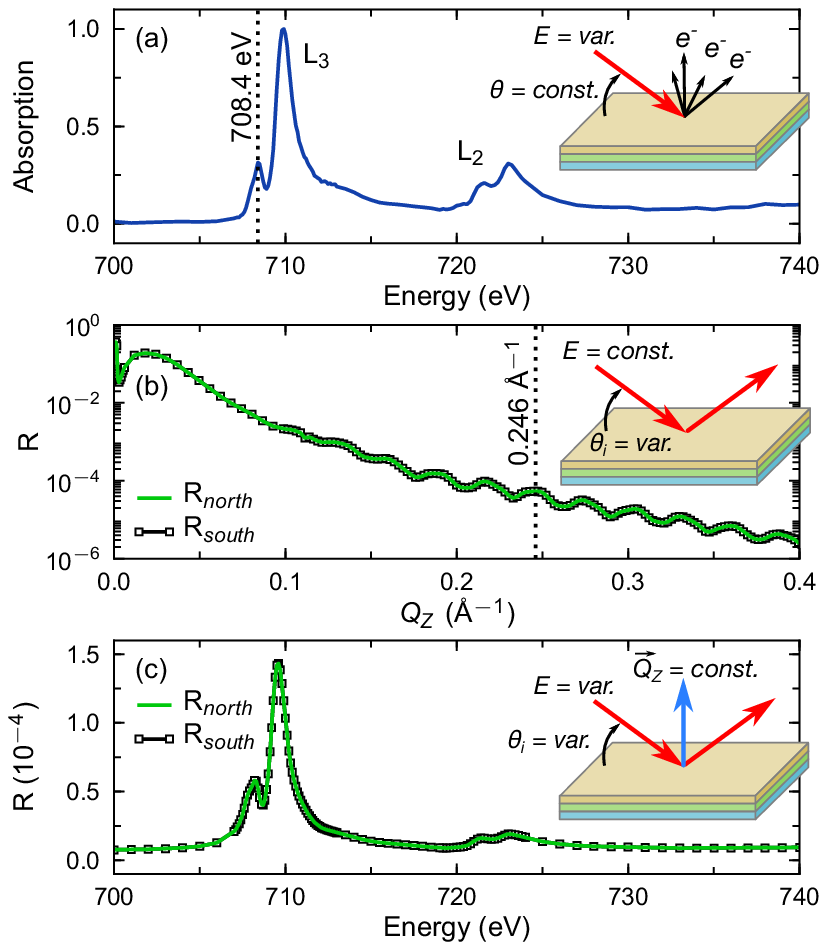}
\caption{
\label{fig:XAS}
(color online)
a) XAS in the vicinity of the Fe L$_{2,3}$ edges for unpolarized X-rays.
b) XRMR curves (parallel and antiparallel) measured at an energy of E = 708.4 eV
using positive circular polarized X-rays and a flipped external magnetic field of $80\,\mathrm{mT}$.
No visual difference between the two curves is apparent, implying the absence of a sizable magnetic signal.
c) Reflected intensity energy scan at Q$_z$ = 0.246$\,$\AA$^{-1}$, a point of the reflectivity curve that
would show a significant splitting in the presence of FM contribution.}
\end{figure}

\begin{figure}
\includegraphics[width=7.0cm]{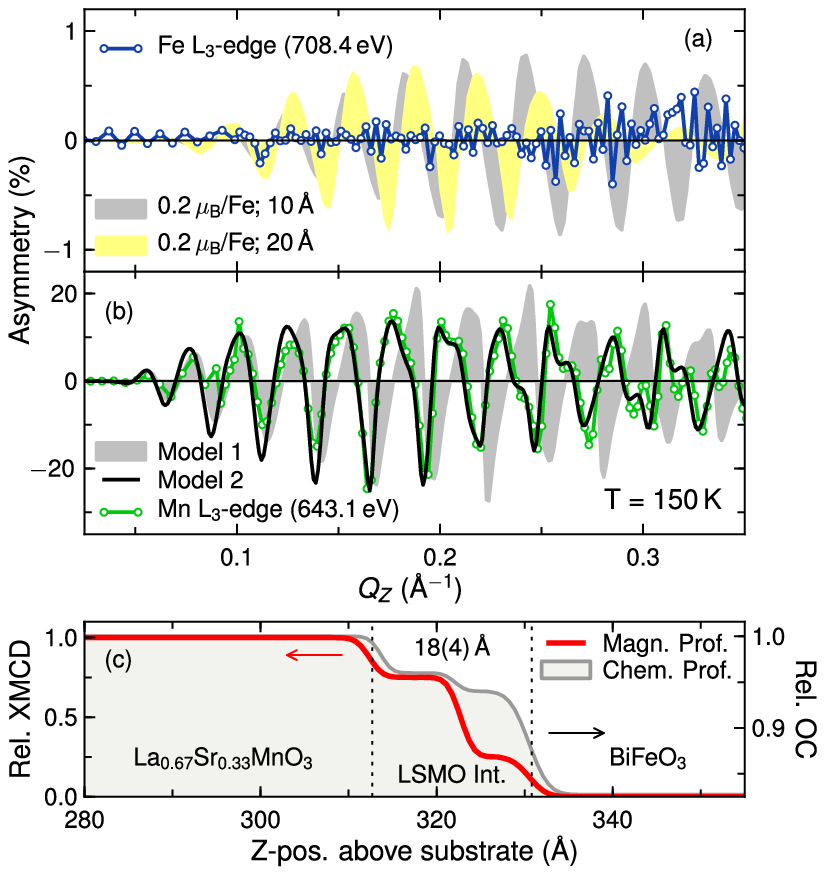}
\caption{
\label{fig:xrmr_asym}
(color online)
a) XRMR magnetic asymmetry signal at the Fe $\mathrm{L_3}$ edge at $708.4\,\mathrm{eV}$ conclusively indicates
no Fe ferromagetic signal. The shaded areas show simulated asymmetries for a hypothetical magnetic moment of
$0.2\,\mathrm{\mu_B/Fe}$ in the first 10$\,$\AA~(grey area) and 20$\,$\AA~(yellow area).
b) XRMR magnetic asymmetry measured at the Mn $\mathrm{L_3}$ edge at $643.1\,\mathrm{eV}$. The best agreement is obtained
for a model with a LSMO interfacial layer with reduced magnetization (black line). For comparison, a homogenous magnetized
LSMO layer does not reproduce the periodicity for smaller $Q_z$ (grey area).
c) Resulting magnetic depth profile. The relative optical constant obtained from resonant XRR, which indicates the change
in the chemical profile, is shown as well.}
\end{figure}

In order to firmly determine whether the ferromagnetic diluted interface layer is located within the LSMO, the BFO, or
across the interface, we have performed element specific X-ray resonant magnetic reflection (XRMR) measurements at the Fe
$\mathrm{L}_{3}$ and Mn $\mathrm{L}_{3}$ edges. XRMR is an extension of standard reflectometry with the use of X-ray
magnetic circular dichroism (XMCD)~\cite{Brue08}. Measurements were performed at the UE56/2-PGM1 beamline at the BESSY II
synchrotron of the Helmholtz Center Berlin, Germany at the ERNSt endstation of the Max Planck Institute for Intelligent
Systems, Stuttgart, Germany~\cite{Brue08}. A representative Fe $\mathrm{L}_{2,3}$ X-ray absorption spectrum (XAS) of the
sample in total electron yield (TEY) is shown in Fig.~\ref{fig:XAS} together with XRMR curves taken at the Fe $\mathrm{L_{3}}$
edge ($\mathrm{E}=708.4\,\mathrm{eV}$) for parallel and antiparallel alignment of the magnetization vector of the sample and
the X-ray beam polarization vector. The XRMR curves in Fig.~\ref{fig:XAS}b) and c) show that the measurements taken for both
polarizations are quasi-identical. This already indicates the absence of any sizable ferromagnetic moment in the BFO layer.
Figure \ref{fig:xrmr_asym}a) shows the magnetic asymmetry ratio of the $Q_z$ measurement taken at the Fe $\mathrm{L_{3}}$
edge (from the data of Fig.~\ref{fig:XAS}b). Simulations of hypothetical FM signals in BFO are shown for comparison
(program ReMagX \cite{Mack10}). A magnetic polarization of $0.2\,\mathrm{\mu_B/Fe^{3+}}$ in the first $10\,\mathrm{\AA}$
(grey area) and $20\,\mathrm{\AA}$ (yellow area) of BFO at the LSMO interface were assumed. Using literature data for the
magneto-optical constants \cite{Bea06} and for the scaling \cite{Mart08}, the simulations clearly show that even such a
small moment would result in a sizable magnetic asymmetry. Our measurements do not show any ferromagnetic signal
originating from the Fe-ions in the BFO layer down to a noise level of 0.3$\,$\%, indicating that the Fe magnetic moments
are not larger than $0.04\,\mathrm{\mu_B/Fe^{3+}}$ in the BFO layer.

As the next step, measurements were performed at the Mn $\mathrm{L_3}$ edge at $\mathrm{E} = 643.1\,\mathrm{eV}$
(see Fig.~\ref{fig:xrmr_asym}b). Qualitative XAS was measured via fluorescence emission and used to calibrate literature
reference data for LSMO \cite{Pell97,Free07,Brue11} to obtain the optical and magneto-optical index of refraction for LSMO.
The degree of circular polarization of $90\,\%$ for the present measurements was taken into account when calculating
the magneto-optical part of the index of refraction. Figure \ref{fig:xrmr_asym}b) shows a significant magnetic asymmetry
for the Mn L$_3$ edge, indicating a spontaneous FM polarization. By simulating the Mn magnetic asymmetry we derived the
Mn-specific magnetic depth profile. The magnetic reflectivity calculations were based on a magneto-optical approach to
fit to the measured reflectivity data \cite{Zak91}, where the polarization dependence of the incident X-ray beam upon the
direction of the dielectric susceptibility tensor is taken into account. We again applied Model I (see Fig.~\ref{fig:pnr_nrex_ref})
without an interfacial layer between the LSMO and BFO layers (grey area). In this case the oscillations of the asymmetry
signal cannot be reproduced. The best fit to the data is obtained when an interface layer with a gradually reduced magnetic
moment in the top 18(4)$\,$\AA~of the LSMO layer was introduced (see solid line in Fig.~\ref{fig:xrmr_asym}b). The resulting
magnetic profile is shown in Fig.~\ref{fig:xrmr_asym}c). This is in good agreement to the result obtained by PNR. The combined
PNR and XRMR analysis indicates a depleted FM region extending about $20\,\mathrm{\AA}$ into the LSMO at the bi-layer interface.

Our results are in contrast to previous XMCD measurements on an inverse BFO/LSMO bi-layer, which yielded an induced
FM interface layer with a magnetic moment of $0.6\,\mu_{B}$/Fe$^{3+}$-ion within the BFO \cite{Yu10}. A similar result was
obtained on a $180\,\mathrm{\AA}$ BFO layer with a top layer of $75\,\mathrm{\AA}$ of CoFeB \cite{Bea08}.
PNR measurements on this bi-layer indicated a $20\,\mathrm{\AA}$ interfacial layer with a FM moment of $1\,\mu_{B}$/Fe$^{3+}$
extending into the BFO layer. The induced FM magnetic moment was explained by orbital or spin reconstructions at the
interface.

\begin{figure}
\includegraphics[width=7.0cm]{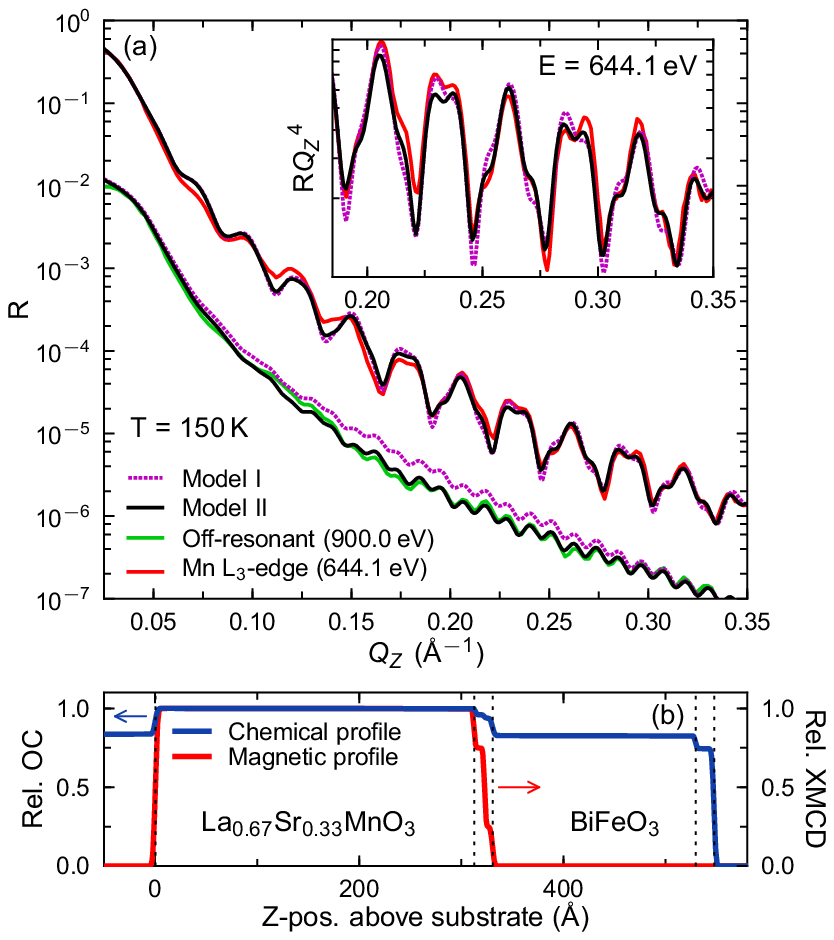}
\caption{
\label{fig:xrr_multifit}
(color online)
a) X-ray reflectivity data measured at 644.1$\,$eV and 900.0$\,$eV, i.e. resonant at the Mn L$_{3}$ edge and off-resonant.
The data was fitted using two different models as described in the text. The inset shows $R Q_z^4$ in the most
sensitive high $Q_z$ region.
b) Resulting chemical (through XRR) and magnetic (PNR and XRMR) depth profiles.}
\end{figure}

In order to elucidate the origin of the reduction of the magnetic moment at the LSMO/BFO interface, we have
performed intrinsic element specific resonant X-ray reflectivity measurements with linear polarization
at the ERNSt endstation at BESSY-II. This did yield valuable information on the local chemistry and
valences at the LSMO/BFO interface. Using the results of the XAS study (see Fig.~\ref{fig:XAS}), XRR curves
were collected at T=150\,K for the Mn L$_3$ edge (maximum XAS signal at 644.1 eV) and off-resonance
at 900 eV. Similarly to the XRMR analysis, simulations of the XRR experimental data were performed with an algorithm we
developed. The XRR curves and the fitted results are shown in Fig.~\ref{fig:xrr_multifit}. It is important to note that
our experiments indicate that the bulk of the LSMO and BFO layers exhibit ideal stoichiometry, and that RMS interfacial
roughnesses between the SrTiO$_3$ substrate, LSMO, and BFO were found to be no larger than 5(1)$\,$\AA. The simulations
did indicate that the XRR data collected in off-resonance condition are highly sensitive to modifications
within the BFO layer. A significant improvement of the theoretical fit to the data was achieved by including a
18(5)$\,$\AA~thick top layer with a 10$\,$\% changed optical constant (black line versus dotted line). This indicates a
modification of the BFO stoichiometry towards impurity phases at the BFO interface to air.

Resonant XRR measurements performed at the Mn $\mathrm{L_3}$ edge offer the opportunity to detect small alterations
in the depth profile of the Mn stoichiometry. The analysis of the Mn XRR data indicated a 17(5)$\,$\AA \ layer inside
the LSMO layer at the interface between LSMO and BFO with a variation of the optical constants (OC) of about 5$\,$\%
(see black solid line in Fig.~\ref{fig:xrr_multifit} in contrast to the purple dotted line for the simpler model assuming
homogeneous layers). As the optical properties vary strongly close to the resonance peak of the Mn $\mathrm{L_3}$ edge,
the absolute change in stoichiometry cannot be precisely quantified. However, the data points towards a modification of
the interface stoichiometry with an altered oxygen and Mn-content in this region \cite{Pell97,Hori05,Pico07}. Even a
slight modification in the oxygen and Mn-concentration would result in a significant deviation of the Mn$^{3+}$/Mn$^{4+}$
ratio. This would have a direct impact on the concentration of the charge carriers which mediate the double exchange
interaction between the ferromagnetically coupled ions. As consequence, the magnetic phase transition temperature and
the ordered magnetic moment decrease \cite{Urus95}. Indeed, an extrapolation of the temperature dependent PNR data indicates
that the interfacial layer has a reduced $T_C$ of $316(10)\,$K (see Supplemental Materials). The thickness of this
chemically altered interfacial layer matches exactly with the region of the reduced magnetic moment as determined independently
by PNR and XRMR. This reduction over 5-6 octahedral sites cannot be explained by epitaxial strain effects alone.
Spin and orbital reconstruction right at the interface will also not fully account for this effect since both would
only take place in the topmost atomic layer.
This indicates that the modified magnetic properties in the interface region between the LSMO and BFO layers is mainly caused
by an altered oxygen and Mn concentration.

In conclusion, we have demonstrated that the combination of complementary reflectometry techniques allowed for the
precise determination of the magnetic and chemical depth profile in LSMO/BFO heterostructures. Resonant element-specific
X-ray reflection measurements indicated an interface region of modified oxygen or Mn-content with the same thickness as
the magnetically diluted interfacial layer within the LSMO film as determined by the magnetic XRMR and PNR measurements.
This magnetically diluted layer hinders the spin polarized current across the interface and hence deteriorates
the functionality of tunneling junctions. Therefore, our result underlines the importance of the precise knowledge
of the chemical composition at the interface. Modifications in the stoichiometry can occur during the growth interruption,
when the LSMO and the BFO layers were completed~\cite{Bea08}. This is a common problem for all transition metal oxide
thin film systems. Previous investigations have shown that the atomic concentrations at the interface can be controlled
systematically by the growth process \cite{Ange04,Leco96,Ohto04}. Changing the deposition conditions would allow for the
engineering of the interface structure and thus enable the growth of artificial heterostructures with specifically desired
properties.

\begin{acknowledgements}
The authors would like to thank D. Manske at the MPI-FKF, Stuttgart, Germany, for fruitful discussions and careful reading of the 
manuscript.
This work was supported by the Australian Research Council through the grant DP110105346 and by AINSE.
V.N. and R.M. acknowledge the support of the ARC (grant LP0991794).
Electron microscopy was carried out at the National Center for Electron Microscopy, Lawrence Berkeley National Laboratory,
supported by the U.S. Department of Energy under Contract No. DE-AC02-05CH11231.
\end{acknowledgements}

\bibliographystyle{apsrev4-1}

\begin{thebibliography}{99}
\bibitem{Yu10}   P.~Yu, $et$ $al.$
                 Phys. Rev. Lett. {\bf 105}, 027201 (2010).
\bibitem{Ohto04} A.~Ohtomo and H.~Y.~Hwang, Nature {\bf 427}, 423 (2004).
\bibitem{Yama04} H.~Yamada, Y.~Ogawa, Y.~Ishii, H.~Sato, M.~Kawasaki, H.~Akoh, and Y.~Tokura, Science {\bf 305}, 646 (2004).
\bibitem{Reyr07} N.~Reyren $et$ $al.$,
                 Science {\bf 317}, 1196 (2007).
\bibitem{Mann10} J.~Mannhart and D.~G.~Schlom, Science {\bf 327}, 1607 (2010).
\bibitem{Hwan12} H.~Y.~Hwang, Y.~Iwasa, M.~Kawasaki, B.~Keimer, N.~Nagaosa, and Y.~Tokura, Nature Mater. {\bf 11}, 103 (2012).
\bibitem{Chak07} J.~Chakhalian, J.~W.~Freeland, H.-U.~Habermeier, G.~Cristiani, G.~Khaliullin, M.~van~Veenendaal,
                 and B.~Keimer, Science {\bf 318}, 1114 (2007).
\bibitem{Driz12} N.~Driza $et$ $al.$,
                 Nature Mater. {\bf 11}, 675 (2012).
\bibitem{Kono04} M.~Konoto, T.~Kohashi, K.~Koike, T.~Arima, Y.~Kaneko, Y.~Tomioka, and Y.~Tokura,
                 Appl. Phys. Lett. {\bf 84}, 2361 (2004).
\bibitem{Cata09} G.~Catalan and J.~F.~Scott, Adv. Mater. {\bf 21}, 2463 (2009).
\bibitem{Zhao06} T.~Zhao $et$ $al.$,
                 Nature Mater. {\bf 5}, 823 (2006).
\bibitem{Bea08a} H.~B\'ea, M.~Gajek, M.~Bibes, and A.~Bart\'el\'emy, J. Phys.: Condens. Matter {\bf 20}, 434221 (2008).
\bibitem{Hamb10} M.~Hambe, A.~Petraru, N.~A.~Pertsev, P.~Munroe, V.~Nagarajan, and H.~Kohlstedt,
                 Adv. Funct. Mater. {\bf 20}, 2436 (2010).
\bibitem{Eere06} W.~Eerenstein, N.~D.~Mathur, and J.~F.~Scott, Nature {\bf 442}, 759 (2006).
\bibitem{Ange04} M.~Angeloni, G.~Balestrino, N.~G.~Boggio, P.~G.~Medaglia, P.~Orgiani, and A.~Tebano,
                 J. Appl. Phys. {\bf 96}, 6387 (2004).
\bibitem{Urus95} A.~Urushibara, Y.~Moritomo, T.~Arima, A.~Asamitsu, G.~Kido, and Y.~Tokura, Phys. Rev. B {\bf 51}, 14103 (1995).
\bibitem{Leco96} P.~Lecoeur, A.~Gupta, P.~R.~Duncombe, G.~Q.~Gong, and G.~Xiao, J. Appl. Phys. {\bf 80}, 513(1996).
\bibitem{Koni99} Y.~Konishi, Z.~Fang, M.~Izumi, T.~Manako, M.~Kasai, H.~Kuwahara, M.~Kawasaki, K.~Terakura1, and Y.~Tokura,
                J. Phys. Soc. Jpn. {\bf 68} 3790 (1999).
\bibitem{Bea08}  H.~B\'ea $et$ $al.$,
                 Phys. Rev. Lett. {\bf 100}, 017204 (2008).
\bibitem{Fitz04} M.~R.~Fitzsimmons $et$ $al.$,
                 J. Magn. Magn. Mater. {\bf 271}, 103 (2004).
\bibitem{Shir69} G.~Shirane and Y.~Yamada, Phys. Rev. {\bf 177}, 858 (1969).
\bibitem{SimulR} ``SimulReflec - Reflectivity Curves Simulations and Fitting," Institute Laue Langevin, (2010).
\bibitem{Brue08} S.~Br\"uck, S.~Bauknecht, B.~Ludescher, E.~Goering, and G.~Sch\"utz,
                 Rev. Sci. Instr. {\bf 79}, 083109 (2008).
\bibitem{Mack10} S.~Macke, S.~Br\"uck, and E.~Goering, ``ReMagX - x-ray magnetic reflectivity tool," (2010).
\bibitem{Bea06}  H.~B\'ea $et$ $al.$,
                 Phys. Rev. B {\bf 74}, 020101 (2006).
\bibitem{Mart08} L.~W.~Martin $et$ $al.$,
                 Nano Lett. {\bf 8}, 2050 (2008).
\bibitem{Pell97} E.~Pellegrin, L.~H.~Tjeng, F.~M.~F.~de Groot, R.~Hesper, G.~A.~Sawatzky, Y.~Moritomo, and Y.~Tokura,
                 J. El. Spec. and Rel. Phenomena {\bf 86}, 115 (1997).
\bibitem{Free07} J.~W.~Freeland $et$ $al.$,
                 J. Phys.: Cond. Matter {\bf 19}, 315210 (2007).
\bibitem{Brue11} S.~Br\"uck $et$ $al.$,
                 New J. Phys. {\bf 13}, 033023 (2011).
\bibitem{Zak91}  J.~Zak, E.~R.~Moog, C.~Liu, and S.~D.~Bader, Phys. Rev. B {\bf 43}, 6423 (1991).
\bibitem{Hori05} K.~Horiba, $et$ $al.$, Phys. Rev. B {\bf 71}, 155420 (2005).
\bibitem{Pico07} S.~Picozzi $et$ $al.$,
                 Phys. Rev. B {\bf 75}, 094418 (2007).
\end{thebibliography}

\end{document}